# Analog of the Carnot engine for fluctuating diffusivity in living cells

Yuichi Itto[1,2]

[1] Science Division, Center for General Education, Aichi Institute of Technology, Aichi 470-0392, Japan

[2] ICP, Universität Stuttgart, 70569 Stuttgart, Germany

**Abstract** Recently, a formal analogy between the fluctuating diffusivity and thermodynamics has been proposed based on phenomena of heterogeneous diffusion observed in living cells. This not only offers the analogs of the quantity of heat and work as well as the internal energy but also achieves that of the Clausius inequality for the entropy concerning diffusivity fluctuations. Here, a discussion is developed about constructing a heat-like engine in terms of the fluctuating diffusivity. The engine constitutes two kinds of processes with the average diffusivity or the average local temperature being kept fixed, along which the fluctuation distribution obeys an exponential law. The efficiency of the engine in a cycle, which quantifies how much the diffusivity change as the analog of work can be extracted, is found to formally coincide with that of Carnot's. During the cycle, the total change of the entropy is also shown to vanish.

# 1. Introduction

The technique of so-called single-particle tracking [1] has established the way of elucidating exotic physical phenomena in a great variety of biological cells (see, e.g., Refs. [2-4]). In particular, the phenomenon of heterogeneous diffusion is at the stage of fundamental importance today (see, e.g., Refs. [5-11] and Refs. [12-14] for recent works and reviews, respectively). A common remarkable feature is that the diffusion coefficient of (macro)molecules widely fluctuates under statistical laws, depending on local regions in the cell at the level of their individual trajectories. The nature of such a fluctuating diffusivity is indispensable for deeper understandings of intracellular diffusion, and accordingly its discovery would be a recent major breakthrough.

Several definite statistical laws of diffusivity fluctuations have been reported/discussed so far, and among them, the exponential law should be distinguished: for a set of individual trajectories, the time-averaged mean square displacement [3,13,14], which is the average of the square displacements obtained from a given trajectory over a certain duration of time, denoted here by $\overline{\Delta^2}$ scales for elapsed time, $\tau$, as $\overline{\Delta^2} \sim D\tau^{\alpha}$, where $D$ stands for the diffusivity, i.e., the diffusion coefficient, and its fluctuation distribution obeys the exponential law written here by

$$P_0(D) \sim \exp\left(-\frac{D}{D_0}\right) \qquad (1)$$

with $D_0$ being the average value of $D \in [0, \infty)$, whereas $\alpha$ is the diffusion exponent taking a certain positive value. Examples are normal diffusion, i.e., $\alpha = 1$, for lipid vesicles in a solution of F-actin filaments [15], transmembrane proteins and lipid molecules on cell membrane [16,17], and the case of $\alpha \neq 1$ for RNA-protein particles in bacteria or yeast cells [18] (see also Ref. [19]), known as anomalous diffusion [20] originated from the viscoelasticity of the cells. There, the statistical fluctuation distribution is of central importance for describing the exponential displacement distribution in the framework of the superposition of the local stochastic processes with respect to it (see, e.g., Refs. [21-23] for relevant works on this framework and also Ref. [24] for the exponential displacement feature). It is noted that the dimension of $D$ changes depending on the value of $\alpha$, which is (approximately) constant, here.

Recently, a formal analogy of the fluctuating diffusivity to thermodynamics has been studied in Ref. [25] based on the experimental results in Ref. [18], and this is precisely the issue we will discuss in a general case of Eq. (1) in the present work. A basic premise there is, for the cell regarded as a medium virtually divided into many small regions or "blocks" for diffusion, that the diffusivity is proportional to temperature of a given block, which varies slowly on a time scale much larger than that of a typical local dynamics such as a random walker, like in the Einstein relation [26] (see Refs. [23,27] and the discussion in Sec. 2 below). It is worth mentioning that such local temperatures have experimentally been measured in living cells, as can be seen, e.g., in Refs. [28-31]. Accordingly, the cell is in nonequilibrium-stationary-state-like situation.

A key in the formal analogy is that at the statistical level, the exponential law in Eq. (1) is formally equivalent to the canonical distribution in Boltzmann-Gibbs statistical mechanics [32] if the diffusivity is identified with the analog of the system energy:

$$P_0(D_i) = \frac{1}{Z}\exp\left(-\frac{D_i}{cT}\right), \qquad Z = \sum_i \exp\left(-\frac{D_i}{cT}\right), \qquad (2)$$

which is purposely expressed in the discrete case for a set of different diffusivities $\{D_i\}_i$ in the local blocks with $D_i$ being the $i$th value of the diffusivity given by $D_i = c\,T_i$ with $c$ and $T_i$ being, respectively, a proportionality factor characterizing mobility and the $i$th value of the local temperature, leading to the average diffusivity written by $D_0 = c\,T$ with $T$ being the average value of the local temperatures in the continuum limit. It is noted that $c$ is considered not to vary over the local blocks since it depends on the diffusion exponent, which is (approximately) constant as mentioned above. As shown in Ref. [33] with the maximum entropy principle [34] (see Ref. [35] for a recent development in this direction) for the entropy associated with diffusivity fluctuations, the identification has also been supported by obtaining a relation analogous to the thermodynamic one concerning temperature in the thermodynamic-like situation, where the average diffusivity can vary slowly and its time scale should be much larger than that of variation of diffusivity fluctuations [see the discussion after Eq. (9) below].

Furthermore, not only the analogs of the quantity of heat and work but also that of the Clausius inequality for the entropy have been established [25]. These analogs are given in terms of the changes of the diffusivity and its statistical fluctuation form in the thermodynamic-like situation, which are considered to be performed by compression/expansion of the cells and the change of temperature, respectively. In this respect, we emphasize the following three points based on relevant experiments. Firstly, it has been shown for their average values that the diffusivity decreases (increases) under compression [36-39] (stretch [37], i.e., expansion) of the cells, whereas the diffusion exponent has almost no change [39]. Secondly, the exponential law in Eq. (1) is expected to be robust against compression or expansion from the fact [37,39] that the statistical property of displacements does not drastically alter. Thirdly, it has been found [36,39] that cell volume/shape is restored by relaxing compression. From these, under Eq. (2), it seems that the diffusivity change is further elaborated at different processes to be realized by controlling compression and expansion as well as temperature, which may constitute a *cycle*.

These observations, therefore, naturally motivate us to examine the possibility for a construction of a "heat engine" to extract the diffusivity change at the statistical level in a cycle consisting of several processes analogous to the thermodynamic ones, along which the exponential law holds. This may be an intriguing problem since the change of the diffusivity plays an essential role, for example, for tuning the rates of biochemical reactions in cells, see, e.g., Ref. [40] (see also Refs. [41,42]).

In this paper, we develop a discussion about constructing an analog of the heat engine in the formal analogy between the fluctuating diffusivity and thermodynamics. This heat-like engine consists of two kinds of processes with the average diffusivity or the average local temperature being kept unchanged, along which the diffusivity obeys the exponential law in Eq. (2). For the diffusivity change as the analog of work, in which the statistical form of the fluctuation distribution is kept fixed [see Eq. (3) in Sec. 2 below], the efficiency of the engine in a cycle, which quantifies how much such a diffusivity change is extractable, is found to formally take that of Carnot's [43,44]. During the cycle, the total change of the entropy associated with diffusivity fluctuations is also shown to vanish.

## 2. Formal analogy between fluctuating diffusivity and thermodynamics

Let us start our discussion with recapitulating the one presented in Ref. [25], which can be developed, without loss of generality, for the case of normal diffusion. Consider a class of heterogeneous diffusion with the exponential diffusivity fluctuation in Eq. (1) in the cells in the case of normal/anomalous diffusion. The cell regarded as the medium imaginarily divided into many local blocks is in a certain state with a set of different diffusivities $\{D_i\}_i$ that follows some statistical fluctuation distribution, $P(D_i)$, which is supposed to slightly deviate from Eq. (1), in general, due to the slow variation of the diffusivity on a long time scale. The relation, $D_i = cT_i$, is then assumed for this set, which gives rise to Eq. (2) in the way mentioned in the Introduction, and accordingly such a state becomes analogous to the "equilibrium state" in the case when $P(D_i)$ is of the form in Eq. (2). We here mention several facts supporting this assumption in the case of anomalous diffusion, in particular. The work in Ref. [18] has suggested the approach of fractional Brownian motion [45], which offers a unified description of both normal and anomalous diffusion, as an underlying stochastic process of RNA-protein particles in bacteria or yeast cells. In Refs. [46,47], based on a generalized Langevin equation characterizing viscoelastic nature of the cell, which yields the diffusion property equivalent to fractional Brownian motion [27], it has been shown, for random walkers such as RNA-protein particles in bacteria, that the mean square displacement is proportional to temperature for large elapsed time, in which the proportionality factor depends on the diffusion exponent through the friction constant. This fact makes it reasonable to suppose that $c$ does not significantly vary in the local blocks since the diffusion exponent is (approximately) constant.

Before proceeding, it may be worth mentioning stochastic processes discussed in Refs.

[22,48], which take into account diffusivity fluctuations in connection with the concept of diffusing diffusivity in Ref. [21], as possible underlying processes in the local blocks. Below, we succinctly explain them in the one dimensional case for the sake of simplicity. Let $x(t)$ be a process of the Brownian motion (but with non-Gaussian displacements) with the time-dependent diffusion coefficient, $D(t)$, described, in the dimensionless form (see Ref. [22] for details), by $dx(t)/dt = \sqrt{2D(t)}\,\xi(t)$ with $\xi(t)$ being the unbiased Gaussian white noise, where $D(t) = Y^2(t)$ is satisfied with $Y(t)$ obeying the following Ornstein-Uhlenbeck process: $dY(t)/dt = -Y(t) + \eta(t)$ with $\eta(t)$ being the unbiased Gaussian white noise independent of $\xi(t)$. As shown in Ref. [22], these processes have yielded normal diffusion with several kinds of the stationary distributions of diffusivity fluctuations, including the one in Eq. (1). Then, in Ref. [48], the former process is generalized based on fractional Brownian motion: $\xi(t)$ is replaced by fractional Gaussian noise, the autocorrelation function of which behaves as a power law. This process with temporal fractionality combined with the latter one has been found to exhibit anomalous diffusion.

Taking into account Eq. (2), we identify the average value of the diffusivity in terms of the fluctuation distribution with the analog of the internal energy: $U_D = \sum_i D_i P(D_i)$. In our present discussion, the average diffusivity can change in the thermodynamic-like situation. There, two states in the medium are distinct each other in the sense that the local fluctuations of the diffusivity in a given state are infinitesimally different from those in the other state. The change of $U_D$ along a process connecting two such states is given by $\delta U_D = \sum_i D_i \delta P(D_i) + \sum_i P(D_i) \delta D_i$, where $\delta P(D_i)$ describes the change of the statistical form of the fluctuation distribution, whereas $\delta D_i$ denotes the change of the diffusivity. From the analogy of the average diffusivity to the internal energy, it is natural to identify the analog of work as

$$\delta' W_D = -\sum_i P(D_i)\delta D_i, \qquad (3)$$

and accordingly the analog of the quantity of heat is identified as

$$\delta' Q_D = \sum_i D_i \,\delta P(D_i). \qquad (4)$$

Therefore, we have the analog of the first law of thermodynamics [32]:

$$\delta' Q_D = \delta U_D + \delta' W_D. \qquad (5)$$

Now, we shall examine the first and second terms in $\delta U_D$ in the case of Eq. (2), which allows us to see relevance of the origins of $\delta P(D_i)$ and $\delta D_i$ to the change of temperature and compression/expansion of the cell. To do so, it is noticed in Eq. (2) that $D_i/(cT)$ does not depend on $c$. Therefore, $\delta P(D_i)$ originates from the change of $T$ while the diffusivity remains unchanged, whereas $\delta D_i$ comes from the change of $c$ with the statistical form of the fluctuation distribution being kept fixed. Regarding the latter, let us recall the point emphasized in the Introduction that the diffusivity decreases (increases) under compression (stretch, i.e., expansion) of the cell on its average. So, it is implied that $c$ reflects a mechanical influence caused by such a compression/expansion process: $c$ plays a role analogous to external parameter. As mentioned in the Introduction, the situation we consider here is the thermodynamic-like one, where the time scale of the change of the average diffusivity due to $c$ and/or $T$ is much larger than that of variation of diffusivity fluctuations. In terms of the compression/expansion process, the existence of this time-scale separation seems to be supported by another point emphasized there that the exponential law may be robust against such a process. The robustness implies that the local property of the fluctuations, which varies (i.e., a short time scale) when the process is performed (i.e., a long time scale), is reorganized in such a way that the statistical property of the fluctuations remains invariant. In addition, it is natural to assume that the time-scale separation is valid within allowed ranges of $c$ and $T$, in which the relation $D_i = cT_i$ as well as the exponential law still holds.

In this way, $\delta P(D_i)$ and $\delta D_i$ are realized, respectively, by the change of temperature and compression/expansion of the cell, here.

The analogs in Eqs. (3) and (4) are thus found to be given by

$$\delta' W_D = -\frac{\partial \langle D_i \rangle}{\partial c} \delta c \tag{6}$$

and

$$\delta' Q_D = \frac{\langle (D_i - \langle D_i \rangle)^2 \rangle}{cT^2} \delta T, \tag{7}$$

where $\delta c$ and $\delta T$ stand for the changes of $c$ and $T$, respectively, and $\langle \bullet \rangle$ denotes the average with respect to $P_0(D_i)$ in Eq. (2). Equation (7) shows that the analog of the quantity of heat is related to the variance of the diffusivity. Equivalently, they are expressed with $Z$ as $\delta' W_D = -T^2 \left( \partial \ln Z / \partial T \right) \delta c$ and $\delta' Q_D = c \left\{ \partial \left[ T^2 \left( \partial \ln Z / \partial T \right) \right] / \partial T \right\} \delta T$.

Next, we see how the analog of the second law of thermodynamics is established in a peculiar manner, (which is based on a basic observation in Ref. [49]). For it, the entropy associated with diffusivity fluctuations is introduced as follows [33]. Recalling that the medium consists of many local blocks, let us imagine a lot of collections constructed by them, where diffusivity fluctuations in a given collection are statistically equivalent to that in the other collections but are locally not. It is reasonable to suppose that the local blocks in each collection are independent each other in terms of the diffusivity since the time-averaged mean square displacement is obtained for individual trajectory. Therefore, the entropy as a measure about uncertainty of the local property of diffusivity fluctuations is found to be given, in a manner similar to the one for deriving the Shannon entropy [34], by $S[P] = -\sum_i P(D_i) \ln P(D_i)$. Below, the exponential law in Eq. (1) (in the discrete case) turns out to be derived by the maximum entropy principle with $S$. From the slow variation of diffusivity fluctuations compared to the typical local dynamics, i.e., a large time-scale separation, the statistical fluctuation distribution to be observed is considered to be the one that may maximize $S$. So, maximization is performed under the constraint on the expectation value of the diffusivity as well as the normalization condition:

$(\delta / \delta P(D_i)) \left\{ S[P] - \kappa \left( \sum_i P(D_i) - 1 \right) - \lambda \left( \sum_i D_i P(D_i) - \bar{D} \right) \right\} = 0,$ where $\kappa$ and $\lambda$ are the Lagrange multipliers concerning the normalization condition and the constraint on the expectation value, giving rise to the following stationary solution, $\hat{P}(D_i) \propto \exp(-\lambda D_i)$, which becomes the exponential law after the choice, $\lambda = 1/D_0$, is made.

With $S$ for the case of the exponential law, we evaluate the change of the following entropy:

$$\tilde{S} = c\, S. \qquad (8)$$

It is noted that the maximum entropy principle with $\tilde{S}$ also offers the exponential diffusivity fluctuation after the redefinition of the Lagrange multipliers. Using the normalization condition on the distribution, the change is calculated to be $\delta \tilde{S} = -c \sum_i \left( \ln P(D_i) \right) \delta P(D_i) + S\, \delta c = \delta' Q_D / T + S\, \delta c,$ where Eq. (2) has been employed at the second equality. Thus, in the case when $c$ is fixed, we have the following relation:

$$\delta \tilde{S} = \frac{\delta' Q_D}{T}, \qquad (9)$$

which not only shows the analogy of $\tilde{S}$ to the thermodynamic entropy but also justifies the identifications in Eqs. (3) and (4). The former can also be seen as follows. In the continuum limit, the entropy with Eq. (1) is calculated as $S = 1 + \ln D_0$. There, it is understood from the dimensional reason of $P_0(D)$ that the entropy is determined up to an additive constant, which cancels if the entropy change is considered. Thus, we have the analog of the thermodynamic relation concerning temperature: $\partial \tilde{S}/\partial D_0 = 1/T$, where the derivative implies that $c$ is kept fixed, supporting its role analogous to external parameter.

Subsequently, let us examine the behavior of the entropy change in the case when $P(D_i)$ differs from $P_0(D_i)$. To quantify the difference between them, we employ the Kullback-Leibler relative entropy [50] given by $K[P \| P_0] = \sum_i P(D_i) \ln [P(D_i)/P_0(D_i)]$, which is positive semidefinite and vanishes if and only if $P(D_i) = P_0(D_i)$. With some fluctuation distribution, $P^*(D_i)$, which satisfies $K[P^* \| P_0] \leq K[P \| P_0]$, we shall describe the change of the form of the fluctuation distribution by $\delta P(D_i) = \{\gamma P^*(D_i) + (1-\gamma) P(D_i)\} - P(D_i)$, where $\gamma$ is a constant in the range $0 < \gamma < 1$. Under this, the change of the relative entropy is written by $\delta_P K[P \| P_0] = K[\gamma P^* + (1-\gamma) P \| P_0] - K[P \| P_0]$, where $\delta_P$ denotes the change with respect to $P(D_i)$. The change is found to be not positive: $\delta_P K[P \| P_0] \leq \gamma \{K[P^* \| P_0] - K[P \| P_0]\} \leq 0$, where the convexity of the relative entropy, i.e., $K[\gamma P^* + (1-\gamma) P \| P_0] \leq \gamma K[P^* \| P_0] + (1-\gamma) K[P \| P_0]$, has been used. The change itself is also given, under the normalization condition on $P(D_i)$, by $\delta_P K[P \| P_0] = -\delta S[P] + \delta' Q_D / D_0$. From these, therefore, in the case when $c$ is fixed, we immediately obtain the analog of the Clausius inequality:

$$\delta \tilde{S} \geq \frac{\delta' Q_D}{T}. \tag{10}$$

Closing this section, we point out the following. In Sec. 4, during the cycle of the analog of the heat engine concerning the fluctuating diffusivity, which we will discuss in the next section, the total change of the entropy turns out to vanish, where $c$ in $\delta \tilde{S}$ is free from condition of fixing its value.

## 3. Analog of the Carnot engine

Based on the formal analogy in the above, here let us construct a heat-like engine for the fluctuating diffusivity with the exponential law in Eq. (2) in the thermodynamic-like situation. As mentioned in the Introduction, we are interested in an efficiency of this engine in a cycle, which tells us how much the diffusivity change as the analog of work in Eq. (6) is extracted. So, in analogy with thermodynamics, we first define the analog of pressure as $p_D = -\partial \langle D_i \rangle / \partial c$, which is given, in the continuum limit, by $p_D = -T$. Then, taking into account the experimental fact [36-39] that the diffusivity decreases (increases) under compression (expansion) at the statistical level, we here suppose, from $D_0 = cT$, that the effect of compression (expansion) is interpreted as the decrease (increase) of the value of $c$. This point is correctly reflected in Eq. (5): given a constant value of $T$, for which $\delta' Q_D = 0$ from Eq. (7), $\delta U_D$ becomes negative (positive) when $c$ decreases, i.e., $\delta'W_D > 0$ (increases, i.e., $\delta'W_D < 0$) since $p_D$ is negative.

On the other hand, $\delta'Q_D$ becomes positive (negative) when $c$ decreases (increases) under $\delta U_D = 0$.

Therefore, as depicted in the plane of the factor $c$ and average local temperature $T$ in Fig. 1, we consider a cycle $A \to B \to C \to D \to A$ consisting of the following four processes. In the processes $A \to B$ and $C \to D$, $D_0$ is kept unchanged, for which $\delta U_D = 0$, and its values are a large diffusivity, $D_L$, and a small diffusivity, $D_S$, respectively, for each of which $T \propto 1/c$. In the processes $B \to C$ and $D \to A$, $T$ is kept fixed at high temperature, $T_H$, and low temperature, $T_C$, respectively, for which $\delta'Q_D = 0$. Then, the values of $c$ at A, B, C, and D are, respectively, $c_1$, $c_2$, $c_3$, and $c_4$. Thus, the following equalities

$$c_1 T_C = c_2 T_H = D_L, \tag{11}$$

$$c_3 T_H = c_4 T_C = D_S, \tag{12}$$

should be fulfilled, from which the following relation holds:

$$\frac{c_1}{c_2} = \frac{c_4}{c_3}, \tag{13}$$

where the ratios appearing are larger than unity since $T_H > T_C$.

During the process $A \to B$, the analog of the quantity of heat increased in the medium is calculated from Eq. (5) with Eq. (6) as $Q_D = \int_{(A)}^{(B)} d'Q_D = -\int_{c_1}^{c_2} dc\, (D_L/c) = D_L \ln(c_1/c_2)$, which is equal to the analog of work, $W_{AB}$, in the process. The analogs of work during other processes are also calculated as follows: $W_{BC} = \int_{(B)}^{(C)} d'W_D = -\int_{c_2}^{c_3} dc\, T_H = D_L - D_S$, $W_{CD} = \int_{(C)}^{(D)} d'W_D = -\int_{c_3}^{c_4} dc\, (D_S/c) = -D_S \ln(c_4/c_3)$, and $W_{DA} = \int_{(D)}^{(A)} d'W_D = -\int_{c_4}^{c_1} dc\, T_C = D_S - D_L$, where Eqs. (11) and (12) have been used at the third equality in $W_{BC}$ and $W_{DA}$. Accordingly, the analog of the total work extracted during the cycle is given by $W_D = W_{AB} + W_{BC} + W_{CD} + W_{DA} = (1 - D_S/D_L) Q_D$, where Eq. (13) has been used at the second equality.

Thus, we obtain the efficiency of the present heat-like engine in the cycle, which is defined by $\eta_D = W_D / Q_D$, as follows:

$$\eta_D = 1 - \frac{D_S}{D_L}, \quad (14)$$

which formally coincides with that of Carnot's: *it depends only on the average diffusivity that remains constant during the process involving the increase/decrease of the analog of the quantity of heat.* (Recall that, in the Carnot engine in thermodynamics, the temperature remains constant during the isothermal process involving the absorption/rejection of the quantity of heat.) Furthermore, we see a remarkable fact that in the case when $c_2 = c_4$ is satisfied, from which $c_1 c_3 = c_2^2$ follows, the efficiency becomes reduced to $\eta_D = 1 - T_C / T_H$, highlighting a striking similarity between the Carnot engine and the present one.

We again emphasize relevant experimental results [36-39] mentioned in the Introduction. From them, the following three points are expected to be realized: the decrease (increase) of the diffusivity under compression (expansion) of the cells, robustness of the exponential diffusivity fluctuation against compression/expansion, and restoration of cell volume/shape by relaxing compression. These seem to be of fundamental for constructing the present Carnot-like engine.

4. **Total change of entropy during the cycle**

As mentioned in Sec. 2, the medium with the exponential diffusivity fluctuation is in a state analogous to the "equilibrium state". Therefore, it is natural to expect that the total change of the entropy in Eq. (8) vanishes in the cycle of the present Carnot-like engine. Indeed, this is the case, as seen below.

The entropy change in the continuum limit is given by $\delta \tilde{S} = \delta' Q_D / T + \left( 1 + \ln D_0 \right) \delta c.$ It is again understood that the entropy in the second term, i.e., the quantity inside the brackets, is determined up to the additive constant mentioned in Sec. 2, the term with which turns out to vanish during the cycle. With this, the entropy change during the four processes are calculated as follows:

$$\tilde{S}_{AB} = \int_{(A)}^{(B)} d\tilde{S} = \int_{c_1}^{c_2} dc \ln D_L = (c_2 - c_1) \ln D_L, \quad \tilde{S}_{BC} = \int_{(B)}^{(C)} d\tilde{S} = \int_{c_2}^{c_3} dc \left[ 1 + \ln(cT_H) \right] =$$

$$(c_3 - c_2) \ln T_H + c_3 \ln c_3 - c_2 \ln c_2, \quad \tilde{S}_{CD} = \int_{(C)}^{(D)} d\tilde{S} = \int_{c_3}^{c_4} dc \ln D_S = (c_4 - c_3) \ln D_S, \text{ and}$$

$$\tilde{S}_{DA} = \int_{(D)}^{(A)} d\tilde{S} = \int_{c_4}^{c_1} dc \left[ 1 + \ln(cT_C) \right] = (c_1 - c_4) \ln T_C + c_1 \ln c_1 - c_4 \ln c_4, \text{ where Eq. (5)}$$

with Eq. (6) in the continuum limit has been used at the second equality in $\tilde{S}_{AB}$ and $\tilde{S}_{CD}$. Using Eqs. (11) and (12), therefore, we find that the total change of the entropy during the cycle vanishes: $\tilde{S}_{total} = \tilde{S}_{AB} + \tilde{S}_{BC} + \tilde{S}_{CD} + \tilde{S}_{DA} = 0.$

From this observation, one may consider that the process is formally "reversible". It should be recalled here that the entropy in Eq. (8) is analogous to the thermodynamic entropy but is associated with diffusivity fluctuations, as mentioned in Sec. 2. This means that during the process, the living cells, which are out of equilibrium and thus the entropy production is relevant, can be seen as those in "equilibrium" with respect to diffusivity fluctuations, which reinforces the statement in the above, showing how the exponential law in Eq. (2) plays a pivotal role. In this context, we wish to mention the following. In Ref. [33], it has been shown that the entropy production rate (in the sense of diffusivity fluctuations) becomes manifestly positive under the mechanism of diffusing diffusivity [21], which admits the exponential law as a stationary solution of the evolution equation of diffusivity fluctuations and is widely employed in complex biological systems. Therefore, further studies along this line may give new insights into "reversibility" of the process.

## 5. Concluding remarks

We have developed a discussion about constructing an analog of the heat engine within the formal analogy between the fluctuating diffusivity and thermodynamics, in which not only the analogs of the quantity of heat and work as well as the internal energy but also that of the Clausius inequality for the entropy associated with diffusivity fluctuations have been established for normal and anomalous diffusion with the exponential law of the fluctuations in living cells. This heat-like engine consists of two kinds of processes to be realized by compression/expansion of the cell and the change of temperature, along which the value of the average diffusivity or average local temperature is kept fixed. To characterize how much the diffusivity change as the analog of work is extracted during a cycle, we have examined the efficiency of the engine and found that it formally coincides with that of Carnot's. We have also seen that the total change of the entropy in the cycle vanishes.

A stretched exponential form of the displacement distribution has been discussed in Ref. [51] for diffusion of phospholipids and proteins in crowded membranes, and such a displacement feature is known to be relevant to a stretched exponential law of diffusivity fluctuations [22] (see also Ref. [52] for such fluctuations of a Weibull type in the context of the diffusion of cells). Recently, in Ref. [53], foundations of statistical mechanics for treating systems with unstable interactions have been introduced and, quite interestingly, a stretched exponential distribution has been proposed as a generalized canonical distribution for describing their equilibrium states. This distribution is characterized by a single exponent determined by a system Hamiltonian and offers the ordinary canonical distribution in a special case, where the exponent is unity, while the conventional thermodynamic relations hold. For the stretched-exponential diffusivity fluctuation, therefore, it seems natural to expect that the formal analogy is formulated with the dependence of all the relevant analogs on the exponent of a similar kind, reproducing the present one in the special case. Therefore, it is of extreme interest to examine if the formal analogy can be further elaborated for the stretched-exponential diffusivity fluctuation.

We also point out a possible relevance of the formal analogy to the diffusivity fluctuation distribution decaying as a power law found in recent experimental works, e.g., in Refs. [54,55], where it has been reported that the diffusion exponent also fluctuates in a wide range. For the formal analogy to be applicable [apart from the similarity with the (generalized) canonical distribution], it seems necessary to incorporate the statistical property of diffusion-exponent fluctuations in the present discussion in an appropriate manner. Cleary, further studies are needed to clarify this point.

**Acknowledgements**

The author would like to thank Rudolf Hilfer for his interest in the present work and for drawing his attention to the work in Ref. [53]. He also thanks I. Iyyappan for pointing out the work in Ref. [56], where fluctuating mobility has been studied in the context of Brownian heat engine [57]. This work has been completed while he has stayed at the Institut für Computerphysik, Universität Stuttgart. He would like to thank the Institut für Computerphysik for hospitality. This work was supported by a Grant-in-Aid for Scientific Research from the Japan Society for the Promotion of Science (No. 21K03394).

**Author contribution**

The author designed research and wrote the paper.

**Data availability**

This paper has no additional data.

**Code availability**

This paper has no code.

Fig. 1

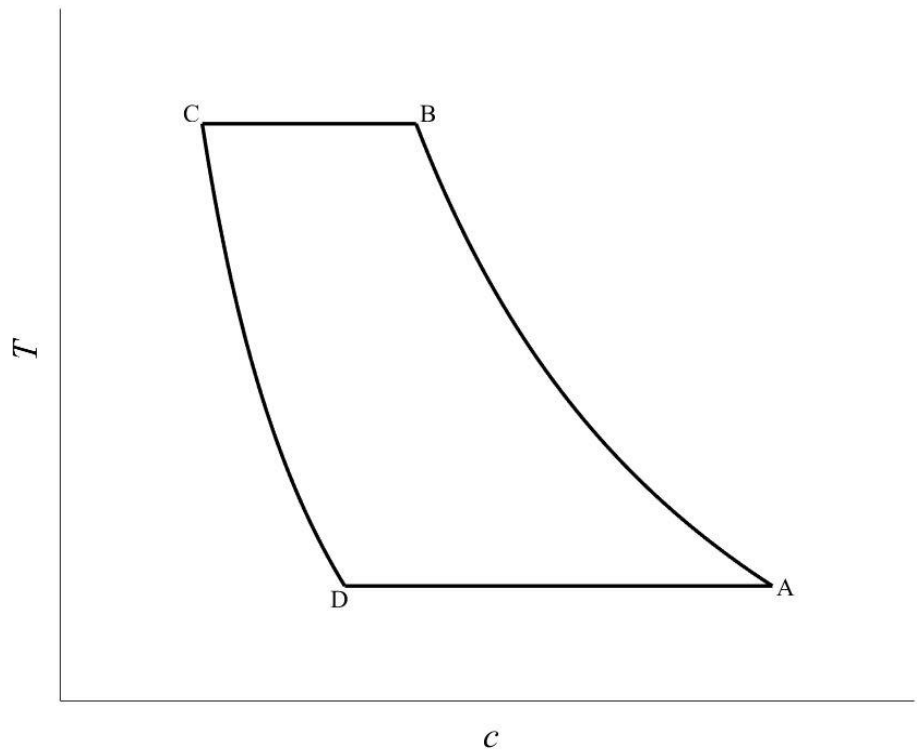

# Figure Caption

Fig. 1

The cycle of the analog of the Carnot engine in the plane of the factor $c$ and average local temperature $T$. The value of $D_0$ is kept unchanged during the processes $A \to B$ and $C \to D$, in which Eqs. (11) and (12) hold, respectively. The value of $T$ remains constant during the processes $B \to C$ and $D \to A$.